\documentclass[
  aps,
  prd,
  reprint,
  superscriptaddress,
  amsmath,
  amssymb,
  nofootinbib,
  longbibliography
]{revtex4-2}

\usepackage[T1]{fontenc}
\usepackage{bm}
\usepackage{graphicx}
\usepackage{tikz}
\usepackage{microtype}
\usepackage{xcolor}
\usepackage[colorlinks=true,allcolors=blue]{hyperref}

\graphicspath{{Figures/}}

\newcommand{\dd}{\mathop{}\!\mathrm{d}}

\begin{document}

\title{Modeling QNMs-Filtered Gravitational Waveforms: \\Approximation Limits and Filtering Effects}

\author{Jun-Xi Shi}
\thanks{Email address: shijunxi23@mails.ucas.ac.cn}
\affiliation{International Centre for Theoretical Physics Asia-Pacific, University of Chinese Academy of Sciences, 100190 Beijing, China}
\affiliation{Dipartimento di Fisica, Sapienza Università 
	di Roma, Piazzale Aldo Moro 5, 00185, Roma, Italy}
\affiliation{INFN, Sezione di Roma, Piazzale Aldo Moro 2, 00185, Roma, Italy}

\author{Zhen-Tao He}
\affiliation{School of Physical Sciences, University of Chinese Academy of Sciences, Beijing 100049, China}

\author{Tingyuan Jiang}
\affiliation{International Centre for Theoretical Physics Asia-Pacific, University of Chinese Academy of Sciences, 100190 Beijing, China}

\author{Yu Tian}
\thanks{Contact author: ytian@ucas.ac.cn}
\affiliation{School of Physical Sciences, University of Chinese Academy of Sciences, Beijing 100049, China}
\affiliation{\textit{Institute of Theoretical Physics, Chinese Academy of Sciences, Beijing 100190, China}}

\author{Dan-Dan Yuan}
\affiliation{International Centre for Theoretical Physics Asia-Pacific, University of Chinese Academy of Sciences, 100190 Beijing, China}

\author{Hongbao Zhang}
\thanks{Contact author: hongbaozhang@bnu.edu.cn}
\affiliation{School of Physics and Astronomy, Beijing Normal University, Beijing 100875, China}
\affiliation{Key Laboratory of Multiscale Spin Physics, Ministry of Education, Beijing Normal University, Beijing 100875, China}

\date{\today}

\begin{abstract}

The waveform left after quasinormal modes filtering of binary gravitational waveforms, often called the 
direct wave, has recently attracted considerable debate. Analytic models of this residual waveform
involve several levels of approximation, while the effects of filtering on the modeling remain 
unclear. These issues require further investigation. In this paper, we examine the approximation 
errors, limits of applicability, and role of filtering in these models. We find that the 
point-particle models examined here show notable discrepancies with the filtered Binary-black-hole waveforms. 
Analytic particle templates based on the stationary-phase approximation perform 
reasonably well for the \((\ell,m)=(2,2)\) multipole, but exhibit substantial limitations for the \((4,4)\). Although the filter is all-pass,
it substantially affects the modeling results. Our analytic calculations on a Schwarzschild 
background further show that vacuum perturbations with Gaussian initial data can also exhibit 
direct-wave-like waveforms after filtering, with their morphology depending on the location of the initial 
data relative to the light ring.
\end{abstract}

\maketitle

\section{Introduction}
Black hole ringdowns are commonly described as superpositions of quasinormal
modes (QNMs), including overtones and nonlinear mode couplings
\cite{BuonannoCookPretorius2007,GieslerEtAl2019Overtones,
GieslerEtAl2025Overtones,Oshita2021Excitation,OshitaCardoso2025,
MitmanEtAl2025,BaibhavEtAl2023,FinchMoore2021,
ZhuEtAl2024Overtones,LiEtAl2022,MaEtAl2021,
MitmanEtAl2023Nonlinear,CheungEtAl2023Nonlinear,
QuadraticGravity1,Shao:2023yjx,
Shi:2026mrr,
Nonlinear6}.
Rational QNM filters~\cite{MaEtAl2022QNMFilters,
MaSunChen2023ModeCleaning,MaSunChen2023Overtone} raise a complementary view:
suppressing a prescribed set of QNM
contributions leaves a decaying, oscillatory residual which has been interpreted
as the direct wave (DW) \cite{OshitaEtAl2025}.

\begin{figure}[!t]
  \centering
  \begingroup
  \def\HierarchyStep{1.25}       
  \def\HierarchyNodePadding{8pt} 
  \def\HierarchyNodeHeight{0.52cm}
  \def\HierarchyArrowWidth{0.95pt}
  \def\HierarchyBoxWidth{0.5pt}
  \newcommand{\DrawHierarchy}[2]{%
  \tikzset{
    hierarchy node/.style={
      font=#1, align=center, draw=black!65,
      line width=\HierarchyBoxWidth,
      minimum height=\HierarchyNodeHeight,
      inner xsep=\HierarchyNodePadding, inner ysep=1pt
    },
    hierarchy arrow/.style={->, >=stealth, line width=\HierarchyArrowWidth},
    hierarchy label/.style={
      font=#2, align=center, anchor=center,
      fill=white, inner xsep=4pt, inner ysep=1.5pt
    }
  }
  \begin{tikzpicture}
    \node[hierarchy node] (bbh) at (0,0)
      {Binary black holes};
    \node[hierarchy node] (particle) at (0,-\HierarchyStep)
      {Linear plunging particle};
    \node[hierarchy node] (nh) at (0,{-2*\HierarchyStep})
      {Near-horizon plunging particle};
    \node[hierarchy node] (dw) at (0,{-3*\HierarchyStep})
      {Direct wave model in Ref.~\cite{OshitaEtAl2025}};
    \node[hierarchy node] (lu) at (0,{-4*\HierarchyStep})
      {Lu model in Ref.~\cite{LuEtAl2026}};

    \draw[hierarchy arrow] (bbh.south) --
      node[hierarchy label, midway] {Point-particle approximation} (particle.north);
    \draw[hierarchy arrow] (particle.south) --
      node[hierarchy label, midway] {Near-horizon approximation} (nh.north);
    \draw[hierarchy arrow] (nh.south) --
      node[hierarchy label, midway] {Stationary-phase approximation} (dw.north);
    \draw[hierarchy arrow] (dw.south) --
      node[hierarchy label, midway] {Simplified potential-barrier screening} (lu.north);
  \end{tikzpicture}
  }
  \DrawHierarchy{\small}{\scriptsize}
  \endgroup
  \caption{Schematic hierarchy of the waveform models and the approximations connecting them.}
  \label{fig:approximation-hierarchy}
\end{figure}
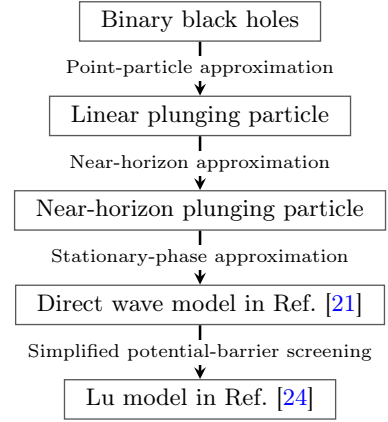

Within the linear Teukolsky framework sourced by a plunging point particle,
Ref.~\cite{OshitaEtAl2025} analytically modeled such residuals using a
near-horizon (NH) expansion \cite{Mino:2008at,Zimmerman:2011dx} and the
stationary-phase approximation (SPA). The model described the filtered
numerical particle waveform reasonably well. Since it is derived in the NH
regime, with its late-time frequency and decay rate governed by the horizon
angular velocity \(\Omega_H\) and surface gravity \(\kappa\) of the background
Kerr black hole, the multi-QNM-filtered residual was interpreted as a non-QNM
component closely related to horizon properties. After further simplifying
the potential-barrier screening in this model, Ref.~\cite{LuEtAl2026} applied
the resulting template to the binary-black-hole (BBH) event GW250114 and
reported evidence of horizon signatures. However, these inferences depend
on both a series of approximations in the waveform model
(see Fig.~\ref{fig:approximation-hierarchy}) and the effects of filtering.

As a result, the physical interpretation of such filtered residual waveforms 
and DW itself quickly became controversial.
Subsequent studies have examined both the physical origin of the filtered
residual and the applicability of its analytic models. Numerical-relativity waveform
tests found that the residual's frequency and damping do not consistently
follow the corresponding horizon-mode predictions
\cite{KankaniMcWilliams2026,DyerEtAl2026}. Beyond horizon-parameter inference,
the filtered waveform has also been proposed as a probe of the NH plunging trajectory
\cite{OshitaEtAl2025,MaWang2026,SunEtAl2026ComplexFrequency}. Results from the time 
domain evolution of a plunging particle, however, showed a small contribution from the region within the light ring
both before and after filtering \cite{Cheung2026LightRing}, while an exact evaluation of
the NH source integral questioned the late-time contribution obtained with
the original SPA treatment
\cite{KubotaEtAl2026Cancellation}; see also Ref.~\cite{KuntzDellaRocca2026Redshift}
for a related conclusion in Schwarzschild spacetime.

Clarifying these issues motivates a systematic analysis of the successive
approximations in Fig.~\ref{fig:approximation-hierarchy} to assess the associated modeling errors. The effect of the
filter itself on waveform modeling also requires attention,
as filtering is central to revealing the DW.
We therefore incorporate filtering into the modeling and examine its effects
in detail. In addition, the $(\ell,m)=(4,4)$ multipole has attracted considerable
attention in recent years because nonlinearity 
and
mode mixing effects are non-negligible in this multipole.
Examining $(4,4)$ can therefore provide further insight into these issues
and test the applicability of the models beyond $(2,2)$.

\begin{figure*}[!t]   
  \includegraphics[width=\textwidth]   
    {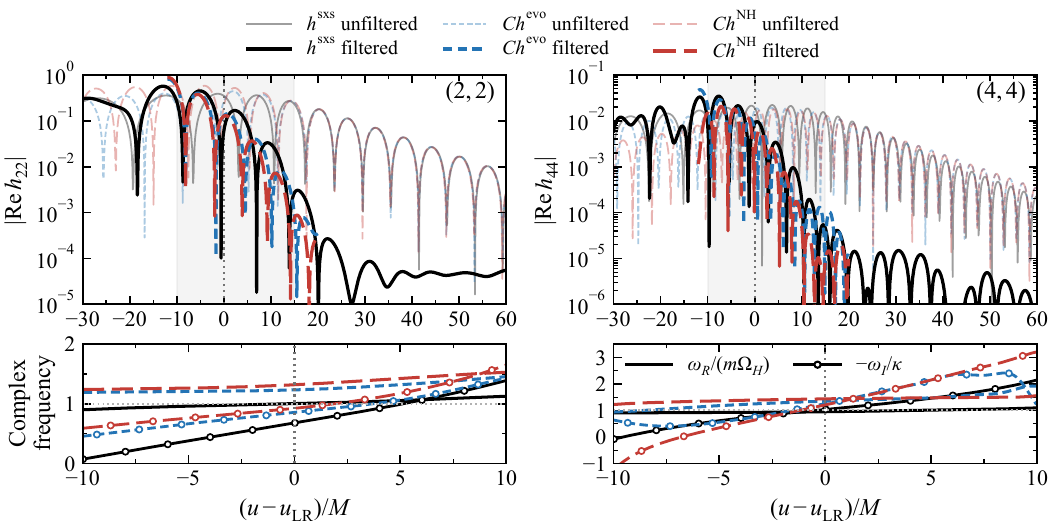}
  \caption{Absolute values of the real parts of the GW strains before and after
  QNM filtering (upper panels), and normalized oscillation frequencies
  and decay rates of the filtered waveforms (lower panels).
  The left and right panels show the $(2,2)$ and $(4,4)$ multipoles, respectively.
  The SXS waveform is taken from SXS:BBH:0305, and both $h^{\rm evo}$ and $h^{\rm NH}$ use its remnant mass and spin.
  The SXS strain \(h^{\rm SXS}\) is shown in black, $h^{\rm evo}$ in blue, and $h^{\rm NH}$ in red;
  each of the latter two is rescaled by a fitted complex constant.
  Line thickness distinguishes unfiltered and filtered waveforms.
  The horizontal axis in each panel shows the retarded time $u$ relative to $u_{\rm LR}$.
  In the lower panels, curves without marks denote the normalized oscillation frequency $\omega_R/(m\Omega_H)$, 
  while curves with circles denote the decay rate $-\omega_I/\kappa$.
  Before filtering, the three waveforms agree well in the ringdown region for both multipoles.
  After filtering, the two particle waveforms $h^{\rm evo}$ and $h^{\rm NH}$ remain roughly consistent with each other,
  but notably differ from $h^{\rm SXS}$.
  }
  \label{fig:compare-SXS-L-NH}
\end{figure*}

\begin{figure*}[!t]   
  \includegraphics[width=\textwidth]   
    {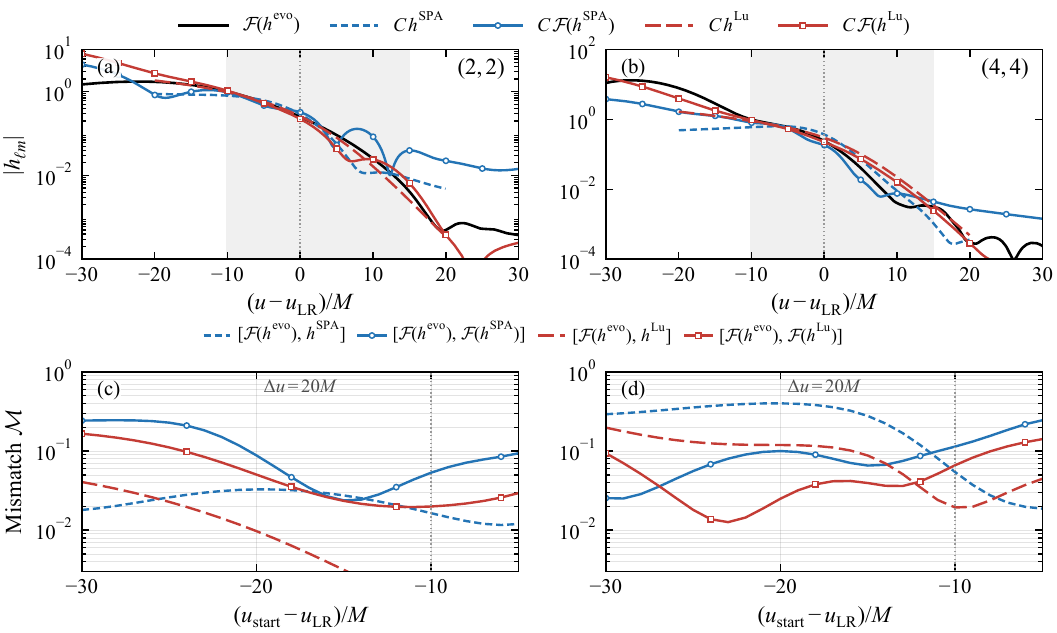}
  \caption{Filtered numerical particle strain \(\mathcal{F}(h^{\rm evo})\) 
  compared with the SPA and Lu waveform models using the SXS:BBH:0305 remnant parameters. Panels (a) and (b) show the
  \((2,2)\) and \((4,4)\) envelopes versus \((u-u_{\rm LR})/M\). Blue and red denote SPA and Lu, respectively, while dashed curves and solid curves with markers
  denote unfiltered models and models processed with the same filter as the numerical waveform, respectively. A constant complex amplitude
  \(C\) is fitted for each waveform model to the filtered numerical waveform over the time window \([-10,15]M\). Panels (c) and
  (d) show mismatches between the complex waveforms in sliding \(20M\) windows as a function of the fit starting time \(u_{\rm start}\). 
  The unfiltered waveform models provide accurate fits in \((2,2)\), but perform dramatically worse in \((4,4)\). The filtering
  generally worsens their agreement with \(\mathcal F(h^{\rm evo})\) in \((2,2)\), but improves it in \((4,4)\) over earlier windows.}
  \label{fig:model-hierarchy}
\end{figure*}

In this paper, we systematically examine these issues.
Section~\ref{sec:Methodology} introduces our methods and the waveforms and
models considered. Section~\ref{sec:model-comparisons} analyzes both the
$(2,2)$ and $(4,4)$ multipoles. In Fig.~\ref{fig:compare-SXS-L-NH}, we first compare BBH waveforms, numerical
particle waveforms, and NH particle waveforms before and after filtering.
In Fig.~\ref{fig:model-hierarchy}, we then focus on the particle models,
examining the approximation errors of the SPA and Lu models and the
non-negligible effects of filtering.
The spin scan in Fig.~\ref{fig:spin_scan} shows that these filtering effects
and the approximation errors in $(4,4)$ persist across a broad spin range,
indicating that the models used in
Refs.~\cite{OshitaEtAl2025,LuEtAl2026} do not adequately describe the
$(4,4)$ residuals considered here. 
In Sec.~\ref{sec:vacuum-case}, we analytically study Schwarzschild
perturbations and find that DW-like waveforms can arise in vacuum
without a continuing external source.
This offers a
possible explanation for the recent disagreements over DWs and further
insight into the role of filtering. We summarize and discuss our results
in Sec.~\ref{sec:discussion}.

Throughout this paper, we set \(c=G=M=1\), where \(M\) is the remnant
black hole mass. The black hole spin parameter is denoted by \(a\) and its
dimensionless spin by \(\chi=a/M\).

\section{Methodology}
\label{sec:Methodology}

We define the filtered residual using the rational multi-QNM filter
\begin{equation}
  F(\omega)
    =\prod_{\ell m n}
      \frac{\omega-\omega_{\ell m n}}{\omega-\omega_{\ell m n}^*},
  \label{eq:multi-qnm-filter}
\end{equation}
where \(\omega_{\ell mn}\) denotes a QNM frequency with angular indices
\((\ell,m)\) and overtone number \(n\). The product runs over the selected
modes, and the asterisk denotes complex conjugation. For a strain multipole
\(h_{\ell m}(u)\) at retarded time \(u\), with Fourier transform
\(\hat h_{\ell m}(\omega)\),
the filtering operation \(\mathcal F\) gives
\begin{equation}
    \mathcal{F}(h_{\ell m})(u)
    =\mathrm{IFT}[
      \hat h_{\ell m}(\omega)F(\omega)],
\end{equation}
where the IFT means inverse Fourier transformation.

The benchmark waveforms in this paper are the numerical BBH gravitational wave (GW) strains \(h^{\rm SXS}\) from SXS catalogs \cite{sxs1,sxs2,sxs3},
and the strains \(h^{\rm evo}\) obtained from linear numerical evolutions of plunging point particles.
 The SXS catalogs used in this paper span a wide dimensionless spin range, 
as shown in Table~\ref{tab:sxssimulations}. The initial condition
of particle plunging is set to an equatorial plunge from the innermost stable circular orbit (ISCO),
throughout this work. 
We specify the trajectory of the point particle using the post-ISCO prescription of Ref.~\cite{Watarai2024} and compute the gravitational waveform by solving the Teukolsky equation~\cite{Teukolsky1973}.

Specifically, we perform the numerical evolution in horizon penetrating hyperboloidally
compactified coordinates, where none of the field characteristics point into the computational domain, and the waveform at future null infinity can be extracted  directly \cite{Anil:2007jw,Ripley:Evolution}.
Exploiting axial symmetry, we factor out the azimuthal dependence and solve the resulting $(2+1)$ dimensional partial differential equation. 
We use sixth-order finite differences in the radial and polar directions and evolve the equation in time using the fourth-order Runge-Kutta method. 
The angular grid excludes the coordinate poles at $\theta=0$ and $\theta=\pi$.
The radially rescaled field at
future null infinity \(\mathcal I^+\) of Weyl scalar \(\Psi_4\) is denoted by
\(\psi_4\equiv\lim\limits_{r\to\infty}r\Psi_4\).
The GW strain \(h\) is obtained from \(\psi_4\) using
fixed-frequency integration (FFI)~\cite{ReisswigPollney2011}.

Applying the NH limit to the linear Teukolsky equation sourced by the point particle,
we obtain the NH waveform \cite{Zimmerman:2011dx}. Specifically,
retaining the full trajectory \((x_p(t),\phi_p(t))\) \cite{OshitaEtAl2025}, we expand the source
to leading order in the NH parameter
\(\epsilon\equiv (r-r_+)/r_+\), which scales as \(e^{2\kappa x}\) near the outer horizon at \(r=r_+\).
Here \(x\) is the tortoise coordinate satisfying
\({\rm d}x/{\rm d}r=(r^2+a^2)/\Delta\), with \(\Delta=r^2-2r+a^2\).
The NH waveform at retarded time \(u\) is
\begin{equation}
  \begin{aligned}
  \psi^{\rm NH}_{4,\ell m}(u) 
&= \int \frac{{\rm d}\omega}{2\pi}
\frac{ e^{-i\omega u}}{2i\omega B_{\ell m\omega}^{\rm inc}}\times\\
\int^{\infty}_{t_0} {\rm d}t\,&e^{i\omega t-im\phi_p}e^{-ik_H x_p}
(\epsilon X_{\ell m\omega}+\mathcal O(\epsilon^2)).
  \end{aligned}
  \label{eq:NHexpandWaveform}
\end{equation}
Here \(k_H=\omega-m\Omega_H\), and \(t_0\) is the start of the source integral.
The frequency-dependent coefficient \(X_{\ell m\omega}\) describes
the leading NH source term, while
\(B^{\rm inc}_{\ell m\omega}\) is the incident coefficient at \(x\to +\infty\) of
the homogeneous ingoing radial solution with unit transmission coefficient.

Direct integration of Eq.~\eqref{eq:NHexpandWaveform}
yields a waveform that includes QNM contributions and agrees well with the
full numerical waveform for \(u>u_{\rm LR}\) (see Fig.~\ref{fig:compare-SXS-L-NH}), where \(u_{\rm LR}\) is the retarded time of the particle's light-ring crossing.
To avoid explicit QNM excitation, we construct the template from the local 
SPA contribution to the leading-order response:
\begin{equation}
  \psi^{\rm SPA}_{4,\ell m}(u) = 
  \frac{X_{\ell m \omega}/(1+\beta)}{2i\omega B_{\ell m\omega}^{\rm inc}}
  \Big|_{\omega = \tilde \omega_m(u)}
  e^{-i\int^u\tilde \omega_m(u')\,{\rm d}u'}.
  \label{eq:SPA-waveform}
\end{equation}
Here \(\beta=-{\rm d}x_p/{\rm d}t\) and \(\Omega={\rm d}\phi_p/{\rm d}t\) are evaluated at a time determined by \(u=t-x_p(t)\),
where the point particle's coordinate time and radial position are directly linked to the retarded time. This is the origin of the ``direct-wave'' picture.
For the leading NH source term,
the SPA yields the following local frequency determined by the particle trajectory:
\begin{equation}
    \tilde \omega_m(u)
    =\frac{m[\beta\Omega_H+\Omega]}{1+\beta}
    -i\frac{2\beta\kappa}{1+\beta}.
  \label{eq:DW-frequency}
\end{equation}
This waveform template retains the full screening prefactor in front of the exponent in Eq.~\eqref{eq:SPA-waveform}. 
To compare with the numerical strain, we project the SPA waveform onto spin-weighted 
spherical harmonics and apply FFI to obtain the
strain multipole \(h^{\rm SPA}_{\ell m}\).

Lu et al. further simplify the screening effect by a factor
 \(\tilde\omega_m(u)-\omega^H_{m}\), yielding a waveform template \cite{LuEtAl2026}:
\begin{equation}
  h^{\rm Lu}_{m}(u) \sim \left[\tilde\omega_m(u)-\omega^H_{m}\right]e^{-i\int^u \tilde \omega_{m}(u')\dd u'},
  \label{eq:DW-general-template}
\end{equation}
where \(\omega^H_{m}=m\Omega_H-i\kappa\) is the corresponding 
horizon frequency. 
For this setting, the template has only the azimuthal dependence.

Previous studies compared unfiltered analytic templates with filtered
waveforms~\cite{OshitaEtAl2025,LuEtAl2026,KankaniMcWilliams2026,DyerEtAl2026,SunEtAl2026ComplexFrequency}.
To assess the effect of this unequal treatment, we apply the same filter and time and
phase corrections to both templates and numerical
waveforms~\cite{MaEtAl2022QNMFilters,LuEtAl2026}.
Unless otherwise stated, we filter QNMs up to a maximum overtone index of 
\(n=5\), with at most one additional fundamental mode to account for 
spherical-spheroidal mixing. 
We also tested the case with a maximum overtone number of \(n=2\), as suggested in Ref.~\cite{SunEtAl2026ComplexFrequency}, but found no qualitative changes in the following results.

The fitting process compares two waveform models in a given time window.
In each analysis window \(W\), we fit
a constant complex amplitude for the waveform template with more approximations. For two waveforms \(a\)
and \(b\), the inner product, fitted amplitude, and normalized mismatch
are defined by
\begin{equation}
  \begin{gathered}
  \langle a,b\rangle_W=\int_W\dd u\,a^*(u)b(u),\qquad
  C_{b\to a}=\frac{\langle b,a\rangle_W}{\langle b,b\rangle_W},\\
  \mathcal M(a,b)=1-
  \mathrm{Re}\left[\frac{\langle a,b\rangle_W}
  {\sqrt{\langle a,a\rangle_W \langle b,b\rangle_W}}\right].
  \end{gathered}
  \label{eq:mismatch}
\end{equation}
In the template comparisons, the mismatch is evaluated after multiplying
each template by its fitted complex amplitude.

\section{Model comparisons }
\label{sec:model-comparisons}

We first compare the SXS waveform with the full particle evolution and the NH
approximation in Fig.~\ref{fig:compare-SXS-L-NH}. The remnant parameters of SXS:BBH:0305 are used
for both particle calculations, with
$\chi_f\simeq0.69$ and the remnant mass set to $M=1$. We align the arrival at
$\mathcal I^+$ of a null ray emitted at the particle's light-ring crossing
with the SXS $(2,2)$ strain peak \cite{OshitaEtAl2025} and rescale each particle waveform by a
fitted complex constant. 

The upper panels compare the absolute values of
the real parts of the strains before and after filtering, with $(2,2)$ on
the left and $(4,4)$ on the right. Time is displayed relative to $u_{\rm LR}$.
For a complex waveform $h(u)$, the real and imaginary parts of its instantaneous complex frequency
are evaluated by
\begin{equation}
\begin{aligned}
  \omega_R(u) &= \mathrm{Re}\!\left[i\,\dot h(u)/h(u)\right],\\
  \omega_I(u) &= \mathrm{Im}\!\left[i\,\dot h(u)/h(u)\right],
\end{aligned}
\end{equation}
respectively. The dot denotes differentiation with respect to $u$. 
The lower panels in Fig.~\ref{fig:compare-SXS-L-NH} show these normalized quantities 
$\omega_R/(m\Omega_H)$ and $-\omega_I/\kappa$ for the filtered waveforms.
Before filtering, the three
waveforms agree well in the ringdown region for both the $(2,2)$ and $(4,4)$
multipoles. After filtering, the two particle waveforms remain roughly
consistent with each other, but differ from the SXS residual in their waveform
profiles and complex-frequency evolution. 

Thus, agreement between unfiltered waveforms in the ringdown region does not imply agreement between their filtered residuals. 
This also raises the question of whether agreement between filtered waveforms near $u_{\rm LR}$ 
reflects similar underlying unfiltered signals. These results reveal a 
significant modeling error when describing filtered BBH waveforms with particle-sourced templates without additional corrections (such as those required in EOB models \cite{EOB-waveform}).
To further examine the approximations underlying the analytic DW templates, we next fix the particle source and compare the SPA and Lu models with the filtered numerical particle waveform.

Figure~\ref{fig:model-hierarchy} compares the SPA and Lu models with the
filtered numerical particle waveform $\mathcal F(h^{\rm evo})$ to assess the errors associated with
successive approximations and the role of filtering in waveform modeling.
For each model, we compare both its unfiltered form and its filtered output under the
same filter used for $\mathcal F(h^{\rm evo})$. The upper panels show the
$(2,2)$ and $(4,4)$ envelopes, with a constant complex amplitude fitted for
each model over $[-10,15]M$. The lower panels show mismatches between the
complex waveforms in sliding $20M$ windows, with the complex amplitude
fitted independently in each window.

Panel (a) of Fig.~\ref{fig:model-hierarchy} gives results broadly consistent
with Ref.~\cite{OshitaEtAl2025}: both the SPA and Lu models describe
$\mathcal F(h^{\rm evo})$ well. Panel (c), however, reveals unexpected behavior.
Within the linear Teukolsky framework with a point-particle source, the SPA
model already incorporates the NH approximation, and the Lu model
introduces further approximations to this SPA model, as discussed in Sec.~\ref{sec:Methodology}. 
Despite these additional
approximations, the Lu model fits $\mathcal F(h^{\rm evo})$ better than the
SPA model. Moreover, since the underlying non-QNM
component of \(\mathcal F(h^{\rm evo})\) is also affected by filtering, one might
expect a model that includes filtering effects to perform better,
yet filtering the models worsens the agreement in this case.

The $(4,4)$ results add to this puzzle. The unfiltered models perform
substantially worse in this multipole, whereas including filtering effects
notably improves the fits over earlier windows. 
This is reversed compared with the results in $(2,2)$.
Since the numerical evolution
is purely linear, these results cannot be attributed to potential nonlinear effects.
These comparisons show that filtering can substantially affect the waveform
and must be taken into account when modeling filtered residuals of GW waveforms.
Figure~\ref{fig:model-hierarchy} also raises concerns about the robustness of
the filtering procedure, similar to those raised in
Refs.~\cite{KankaniMcWilliams2026,Han:2026cgd}. These results suggest limitations in applying 
the models developed in Refs.~\cite{OshitaEtAl2025,LuEtAl2026} to the filtered $(4,4)$ residuals. 
Of course, possible numerical effects associated with the filtering
procedure require further analysis.

\begin{figure}[!t]   
  \includegraphics[width=0.45\textwidth]   
  {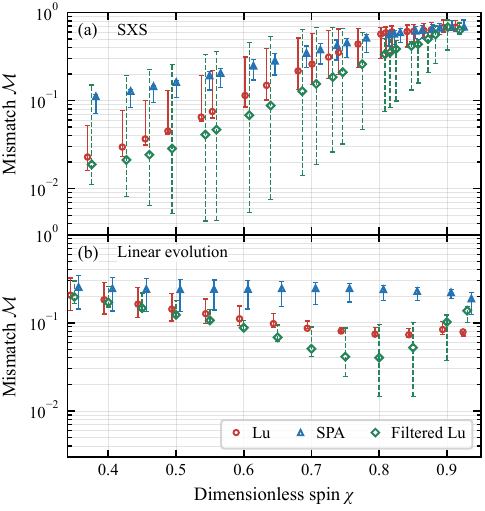}
    \caption{Mismatch of unfiltered \((4,4)\) Lu and SPA models and the filtered Lu model against filtered strain as a function of the
  dimensionless remnant spin \(\chi\). Panels (a) and (b) contain 22 SXS BBH waveforms and
  13 linear evolutions with point-particle sources, respectively. Red circles and blue triangles denote the unfiltered Lu and SPA models,
  respectively, while green diamonds denote the filtered Lu model, processed with the same filter as the corresponding numerical waveform. A constant
  complex amplitude is fitted independently for each model in each \(20M\) window. Markers show the median mismatch over
  windows with start times satisfying \(-20<(u_{\rm start}-u_{\rm LR})/M <-10\), and error bars span the corresponding minimum and maximum.
  Horizontal offsets are for visibility only. }
  \label{fig:spin_scan}
\end{figure}

\begin{table}[!t]
\centering
\caption{SXS simulation IDs and remnant dimensionless spins used in the
spin scan.}
\label{tab:sxssimulations}
\begin{tabular}{cc@{\qquad}cc}
\hline\hline
SXS ID & $\chi_f$ & SXS ID & $\chi_f$ \\
\hline
3976 & 0.376 & 3584 & 0.746 \\
3895 & 0.427 & 2101 & 0.775 \\
2083 & 0.461 & 3921 & 0.809 \\
2089 & 0.494 & 2105 & 0.816 \\
2084 & 0.543 & 3916 & 0.825 \\
3533 & 0.560 & 3980 & 0.847 \\
2328 & 0.608 & 2099 & 0.857 \\
2096 & 0.640 & 1495 & 0.872 \\
2325 & 0.686 & 2512 & 0.883 \\
3926 & 0.707 & 3979 & 0.901 \\
2097 & 0.732 & 2423 & 0.919 \\
\hline\hline
\end{tabular}
\end{table}

We next extend the comparison to a range of remnant spins in
Fig.~\ref{fig:spin_scan}, focusing on the $(4,4)$ multipole. 
The SXS simulations used here are listed in Table~\ref{tab:sxssimulations}. We compare
$h^{\rm SPA}$, $h^{\rm Lu}$, and $\mathcal F(h^{\rm Lu})$ with
$\mathcal F(h^{\rm SXS})$ from 22 SXS BBH simulations spanning
$0.376\leq\chi\leq0.919$ in panel (a), and with $\mathcal F(h^{\rm evo})$
from 13 linear particle evolutions in panel (b). For each remnant, we construct
the models from a particle geodesic on the Kerr background with the corresponding parameters.
The alignment between BBH and particle models follows the same procedure as in Fig.~\ref{fig:compare-SXS-L-NH}.
In each case, the same filter is applied to $h^{\rm Lu}$ and the numerical
waveform, and, for panel (a), the filter also includes the quadratic frequency
$2\omega_{220}$. We evaluate mismatches in sliding $20M$ windows with starting
times between $-20M$ and $-10M$ relative to $u_{\rm LR}$, fitting a constant
complex amplitude independently for each model in each window. The markers
show the median mismatch across these windows, and the error bars span the
minimum and maximum values.

Figure~\ref{fig:spin_scan} shows that the models generally perform worse for
$(4,4)$ over a broad spin range. In panel (a), the fits of all three models
deteriorate as the spin increases, and none adequately describes the filtered
SXS waveforms at high spin. The fits using $\mathcal F(h^{\rm Lu})$ also show
strong sensitivity to the analysis window. In panel (b), the mismatches of
the three models are relatively stable across the spin range, but remain
notably higher than in the $(2,2)$ case. Filtering the Lu model generally
reduces the mismatch in both the SXS and linear evolution cases.

Another counterintuitive feature appears at lower spins: templates derived
from the linear Teukolsky equation with a point-particle source fit the BBH
waveforms better than the particle waveforms. Since the BBH system has richer
dynamics than the point-particle system, this mismatch ordering indicates
that a good fit does not necessarily imply that the underlying dynamics have
been faithfully captured. In the next section, we explore a possible reason
for these apparent inconsistencies in waveform modeling.

\section{Vacuum case}
\label{sec:vacuum-case}
To explore the role of the filter in these modeling discrepancies, we next
examine how it shapes the response in a simpler setting without a continuing
external source. We consider vacuum perturbations of a Schwarzschild black
hole with Gaussian initial data and, following Ref.~\cite{MaWang2026}, examine
the anti-causal-pole (ACP) contribution to the filtered waveform. The polar perturbations satisfy the Zerilli equation
\begin{equation}
  \left[-\partial_t^2+\partial_x^2-V_\ell^{\rm Z}(r)\right]
\Phi_{\ell m}(t,x)=0,
\end{equation}
where we fix \((\ell,m)=(2,2)\) and drop these indices below.
No external source is present for $t>0$, as we prescribe the initial data as
\(\Phi(x)|_{t=0}=0\) and
\(\partial_t\Phi(x)|_{t=0}=10^{-3} \,e^{-(x-x_c)^2/(2\sigma^2)}\) with
center \(x_c\) and width \(\sigma\).
We also evaluate the waveforms at $\mathcal I^+$.

\begin{figure}[!t]   
  \includegraphics[width=0.45\textwidth]   
    {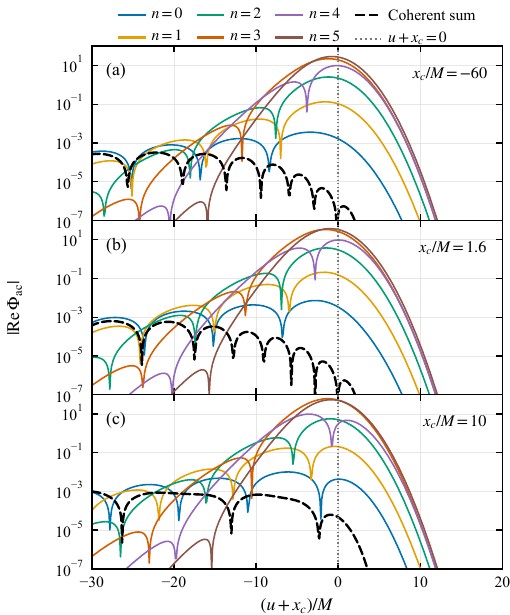}
  \caption{ACP contributions generated by Gaussian velocity initial data for the Schwarzschild
  $(\ell,m)=(2,2)$ Zerilli field, with no external source for \(t>0\). Panels (a)--(c) correspond to Gaussian
  centers \(x_c/M=-60\), \(1.6\), and \(10\), respectively, with fixed amplitude and width \(\sigma/M=2\). 
  The curves labeled \(n=0,\ldots,5\) show the individual ACP contributions; 
  the black dashed curve shows their coherent sum, $\Phi_{\rm ac}$. 
  Coherent cancellation obscures the initial-data dependence of individual 
  pole contributions.
}
  \label{fig:nosourceGaussian}
\end{figure}

Applying the filter to the Zerilli waveform is equivalent to modifying the
Green function (see Eq.~(6) of Ref.~\cite{MaWang2026}). In the pole sector,
the filter cancels the selected QNM poles in the lower half-plane and
introduces ACPs in the upper half-plane. Retaining the ACP contributions
through $n=5$ and taking the inverse Fourier transform gives
\begin{equation}
  \Phi_{\rm ac}(u)
  =-2\,\mathrm{Re}\sum_{n=0}^{5}E_ne^{-i \omega^*_n u}
    \int_{-\infty}^{-u}\dd x\,\mathcal R^{\rm in}_{\omega^*_n}(x)
    \, \partial_t\Phi(x)|_{t=0},
  \label{eq:gaussian-ac-solution}
\end{equation}

where the excitation factor is
\[
E_n=\operatorname*{Res}_{\omega=\omega_n^*}
\frac{F(\omega)/F(-\omega)}
{2\omega\,\mathcal B_\omega^{\rm inc}}.
\]
Here \(\mathcal B_\omega^{\rm inc}\) is the incident coefficient of the
Zerilli in-mode solution at \(x\to+\infty\), with the normalized near horizon asymptotic behavior
\(\mathcal R_\omega^{\rm in}\sim e^{-i\omega x}\). It is distinct from the
Teukolsky coefficient \(B_{\ell m\omega}^{\rm inc}\) used above.
The upper integration limit \(-u\) follows from the anti-causal bound \(u<-x\).

Figure~\ref{fig:nosourceGaussian} shows the individual ACP contributions and
their coherent sum: $\Phi_{\rm ac}$. To examine their sensitivity to the
location of the initial data, we fix the Gaussian amplitude and width and
vary its center from $x_c/M=-60$ through the light-ring value
$x_c/M\simeq1.6$ to $x_c/M=10$. The individual pole contributions change with
the Gaussian center and peak near \(u=-x_c\). Despite these differences,
the summation makes the large individual ACP contributions cancel coherently, leaving a much smaller
sum with a qualitatively similar DW-like oscillatory decay in all three
cases, reminiscent of the particle-driven result in Ref.~\cite{MaWang2026}.

When the Gaussian center lies outside the light ring, the DW-like morphology becomes less pronounced. By contrast, even a Gaussian centered at $x_c/M=-60$ produces a waveform that differs only slightly from the $x_c/M=1.6$ case. These observations suggest that, in our setup, a pronounced DW-like morphology may begin to emerge as the Gaussian center is moved inward toward the light ring, qualitatively similar to the findings of Refs.~\cite{Cheung2026LightRing,DW-lightring}. 

These results show that DW-like waveforms persist even without a continuing external source. The apparent inconsistencies discussed above may therefore arise because filtering does not simply expose a non-QNM component, but also shapes the waveform through the properties of the filter itself. We should therefore examine to what extent the DW morphology and properties reflect the underlying non-QNM component, and to what extent they arise from filtering.

\section{Discussion}
\label{sec:discussion}
In this paper, we examined the approximation errors and applicability of
particle-based models for QNM-filtered waveforms, together with the role of
filtering in their interpretation. Our results reveal limitations of these
models, particularly in the $(4,4)$ multipole, and show that filtering can
substantially affect their agreement with numerical waveforms. The
Schwarzschild calculations further demonstrate that DW-like morphology can
arise without a continuing external source. 
These findings suggest that agreement in waveform shape or a low mismatch between filtered 
waveforms does not necessarily imply similar underlying source dynamics. Inferences based on 
QNM-filtered waveforms must therefore account for the effects of filtering itself.
This paper provides an analysis of recent debates and offers a new perspective on QNM filtering and DWs.

Ringdown waveforms carry important information about strong-field gravity, making a deeper 
analysis of their non-QNM components essential. This calls for careful use of filtering 
\cite{SunEtAl2026ComplexFrequency} or 
the development of alternative analysis methods 
\cite{NakamotoEtAl2026PoleSplitting,WellerEtAl2026Driven}. In future work, we will further explore 
possible applications of filtering, including its use in probing nonlinear effects in general 
relativity.

\begin{acknowledgments}
This work is supported by
the National Natural Science Foundation of China under Grants 
No. 12375058, No. 12361141825, and
No. 12575047.
The author J. S. acknowledges financial support
from the China Scholarship Council (CSC) for joint PhD
training in Rome.
\end{acknowledgments}

\bibliography{references}

@misc{OshitaEtAl2025,
  author        = {Oshita, N. and Ma, S. and Chen, Y. and Yang, H.},
  title         = {Probing Direct Waves in Black Hole Ringdowns},
  year          = {2025},
  eprint        = {2509.09165},
  archiveprefix = {arXiv},
  primaryclass  = {gr-qc},
  doi           = {10.48550/arXiv.2509.09165}
}

@article{BuonannoCookPretorius2007,
   title={Inspiral, merger, and ring-down of equal-mass black-hole binaries},
   volume={75},
   ISSN={1550-2368},
   url={http://dx.doi.org/10.1103/PhysRevD.75.124018},
   DOI={10.1103/physrevd.75.124018},
   number={12},
   journal={Physical Review D},
   publisher={American Physical Society (APS)},
   author={Buonanno, Alessandra and Cook, Gregory B. and Pretorius, Frans},
   year={2007},
   month=June }

@article{GieslerEtAl2019Overtones,
  author = {Giesler, M. and Isi, M. and Scheel, M. A. and Teukolsky, S.},
  title = {Black Hole Ringdown: The Importance of Overtones},
  journal = {Phys. Rev. X},
  volume = {9},
  pages = {041060},
  year = {2019},
  eprint = {1903.08284},
  archiveprefix = {arXiv}
}

@article{GieslerEtAl2025Overtones,
  author = {Giesler, M. and others},
  title = {Overtones and Nonlinearities in Binary Black Hole Ringdowns},
  journal = {Phys. Rev. D},
  volume = {111},
  pages = {084041},
  year = {2025},
  eprint = {2411.11269},
  archiveprefix = {arXiv}
}

@article{Oshita2021Excitation,
  author = {Oshita, N.},
  title = {Ease of Excitation of Black Hole Ringing: Quantifying the Importance of Overtones by the Excitation Factors},
  journal = {Phys. Rev. D},
  volume = {104},
  pages = {124032},
  year = {2021},
  eprint = {2109.09757},
  archiveprefix = {arXiv}
}

@article{OshitaCardoso2025,
    author = "Oshita, Naritaka and Cardoso, Vitor",
    title = "{Reconstruction of ringdown with excitation factors}",
    eprint = "2407.02563",
    archivePrefix = "arXiv",
    primaryClass = "gr-qc",
    reportNumber = "YITP-24-81, RIKEN-iTHEMS-Report-24",
    doi = "10.1103/PhysRevD.111.104043",
    journal = "Phys. Rev. D",
    volume = "111",
    number = "10",
    pages = "104043",
    year = "2025"
}

@article{MitmanEtAl2025,
    author = "Mitman, Keefe and others",
    title = "{Probing the ringdown perturbation in binary black hole coalescences with an improved quasinormal mode extraction algorithm}",
    eprint = "2503.09678",
    archivePrefix = "arXiv",
    primaryClass = "gr-qc",
    doi = "10.1103/qq1g-jlnw",
    journal = "Phys. Rev. D",
    volume = "112",
    number = "6",
    pages = "064016",
    year = "2025"
}

@article{BaibhavEtAl2023,
    author = "Baibhav, Vishal and Cheung, Mark Ho-Yeuk and Berti, Emanuele and Cardoso, Vitor and Carullo, Gregorio and Cotesta, Roberto and Del Pozzo, Walter and Duque, Francisco",
    title = "{Agnostic black hole spectroscopy: Quasinormal mode content of numerical relativity waveforms and limits of validity of linear perturbation theory}",
    eprint = "2302.03050",
    archivePrefix = "arXiv",
    primaryClass = "gr-qc",
    doi = "10.1103/PhysRevD.108.104020",
    journal = "Phys. Rev. D",
    volume = "108",
    number = "10",
    pages = "104020",
    year = "2023"
}

@article{FinchMoore2021,
    author = "Finch, Eliot and Moore, Christopher J.",
    title = "{Modeling the ringdown from precessing black hole binaries}",
    eprint = "2102.07794",
    archivePrefix = "arXiv",
    primaryClass = "gr-qc",
    reportNumber = "LIGO document number P2100036-v2",
    doi = "10.1103/PhysRevD.103.084048",
    journal = "Phys. Rev. D",
    volume = "103",
    number = "8",
    pages = "084048",
    year = "2021"
}

@article{ZhuEtAl2024Overtones,
    author = "Zhu, Hengrui and Ripley, Justin L. and C{\'a}rdenas-Avenda{\~n}o, Alejandro and Pretorius, Frans",
    title = "{Challenges in quasinormal mode extraction: Perspectives from numerical solutions to the Teukolsky equation}",
    eprint = "2309.13204",
    archivePrefix = "arXiv",
    primaryClass = "gr-qc",
    doi = "10.1103/PhysRevD.109.044010",
    journal = "Phys. Rev. D",
    volume = "109",
    number = "4",
    pages = "044010",
    year = "2024"
}

@article{LiEtAl2022,
    author = "Li, Xiang and Sun, Ling and Lo, Rico Ka Lok and Payne, Ethan and Chen, Yanbei",
    title = "{Angular emission patterns of remnant black holes}",
    eprint = "2110.03116",
    archivePrefix = "arXiv",
    primaryClass = "gr-qc",
    doi = "10.1103/PhysRevD.105.024016",
    journal = "Phys. Rev. D",
    volume = "105",
    number = "2",
    pages = "024016",
    year = "2022"
}

@article{MaEtAl2021,
    author = "Ma, Sizheng and Giesler, Matthew and Varma, Vijay and Scheel, Mark A. and Chen, Yanbei",
    title = "{Universal features of gravitational waves emitted by superkick binary black hole systems}",
    eprint = "2107.04890",
    archivePrefix = "arXiv",
    primaryClass = "gr-qc",
    doi = "10.1103/PhysRevD.104.084003",
    journal = "Phys. Rev. D",
    volume = "104",
    number = "8",
    pages = "084003",
    year = "2021"
}

@article{MitmanEtAl2023Nonlinear,
  author = {Mitman, K. and others},
  title = {Nonlinearities in Black Hole Ringdowns},
  journal = {Phys. Rev. Lett.},
  volume = {130},
  pages = {081402},
  year = {2023},
  eprint = {2208.07380},
  archiveprefix = {arXiv}
}

@article{CheungEtAl2023Nonlinear,
  author = {Cheung, M. H.-Y. and others},
  title = {Nonlinear Effects in Black Hole Ringdown},
  journal = {Phys. Rev. Lett.},
  volume = {130},
  pages = {081401},
  year = {2023},
  eprint = {2208.07374},
  archiveprefix = {arXiv}
}

@article{LuEtAl2026,
  author  = {Lu, N. and Ma, S. and Piccinni, O. J. and Chen, Y. and Sun, L.},
  title   = {{GW250114} Reveals Black Hole Horizon Signatures},
  journal = {Nature},
  volume  = {655},
  pages   = {71--75},
  year    = {2026},
  doi     = {10.1038/s41586-026-10696-0}
}

@misc{MaWang2026,
  author        = {Ma, S. and Wang, H.-Y.},
  title         = {Foundations of Direct Waves in Schwarzschild Ringdown},
  year          = {2026},
  eprint        = {2607.06677},
  archiveprefix = {arXiv},
  primaryclass  = {gr-qc},
  doi           = {10.48550/arXiv.2607.06677}
}

@misc{KankaniMcWilliams2026,
  author        = {Kankani, A. and McWilliams, S. T.},
  title         = {The Direct Wave is Not a Meaningful Test of Horizon Properties},
  year          = {2026},
  eprint        = {2607.02380},
  archiveprefix = {arXiv},
  primaryclass  = {gr-qc},
  doi           = {10.48550/arXiv.2607.02380}
}

@misc{Han:2026cgd,
    author = "Han, Wen-Biao and Jiang, Ye",
    title = "{Horizon-redshift transfer in black-hole direct-wave damping}",
    eprint = "2607.08570",
    archivePrefix = "arXiv",
    primaryClass = "gr-qc",
    month = "7",
    year = "2026"
}

@article{MaEtAl2022QNMFilters,
  author        = {Ma, Sizheng and Mitman, Keefe and Sun, Ling and Deppe, Nils and H\'ebert, Fran\c{c}ois and Kidder, Lawrence E. and Moxon, Jordan and Throwe, William and Vu, Nils L. and Chen, Yanbei},
  title         = {Quasinormal-mode filters: A new approach to analyze the gravitational-wave ringdown of binary black-hole mergers},
  journal       = {Physical Review D},
  volume        = {106},
  number        = {8},
  pages         = {084036},
  year          = {2022},
  doi           = {10.1103/PhysRevD.106.084036},
  eprint        = {2207.10870},
  archiveprefix = {arXiv},
  primaryclass  = {gr-qc}
}

@article{MaSunChen2023ModeCleaning,
  author        = {Ma, Sizheng and Sun, Ling and Chen, Yanbei},
  title         = {Black Hole Spectroscopy by Mode Cleaning},
  journal       = {Physical Review Letters},
  volume        = {130},
  number        = {14},
  pages         = {141401},
  year          = {2023},
  doi           = {10.1103/PhysRevLett.130.141401},
  eprint        = {2301.06705},
  archiveprefix = {arXiv},
  primaryclass  = {gr-qc}
}

@article{MaSunChen2023Overtone,
  author        = {Ma, Sizheng and Sun, Ling and Chen, Yanbei},
  title         = {Using Rational Filters to Uncover the First Ringdown Overtone in {GW150914}},
  journal       = {Physical Review D},
  volume        = {107},
  number        = {8},
  pages         = {084010},
  year          = {2023},
  doi           = {10.1103/PhysRevD.107.084010},
  eprint        = {2301.06639},
  archiveprefix = {arXiv},
  primaryclass  = {gr-qc}
}

@misc{DyerEtAl2026,
  author        = {Dyer, Richard and Chung, Adrian Ka-Wai and Moore, Christopher J.},
  title         = {Modeling Direct Waves in Binary Black Hole Ringdowns},
  year          = {2026},
  eprint        = {2606.25021},
  archiveprefix = {arXiv},
  primaryclass  = {gr-qc},
  doi           = {10.48550/arXiv.2606.25021}
}

@misc{Cheung2026LightRing,
  author        = {Cheung, Mark Ho-Yeuk},
  title         = {How Much of the Black Hole Ringdown Is Sourced within the Light Ring?},
  year          = {2026},
  eprint        = {2608.29466},
  archiveprefix = {arXiv},
  primaryclass  = {gr-qc},
  doi           = {10.48550/arXiv.2608.29466}
}

@misc{SunEtAl2026ComplexFrequency,
  author        = {Sun, Ling and Ma, Sizheng and Chen, Patrick P. and Laeuger, Andrew and Lu, Neil and Chen, Yanbei},
  title         = {Complex Frequency Evolution of Direct Waves from Binary Black Hole Mergers},
  year          = {2026},
  eprint        = {2608.23209},
  archiveprefix = {arXiv},
  primaryclass  = {gr-qc},
  doi           = {10.48550/arXiv.2608.23209}
}

@misc{KubotaEtAl2026Cancellation,
  author        = {Kubota, Kei-ichiro and Manita, Yusuke and Motohashi, Hayato and Watarai, Daiki},
  title         = {Exact Cancellation of Horizon Modes in {Kerr} Ringdown},
  year          = {2026},
  eprint        = {2609.07326},
  archiveprefix = {arXiv},
  primaryclass  = {gr-qc},
  doi           = {10.48550/arXiv.2609.07326}
}

@misc{WellerEtAl2026Driven,
  author        = {Weller, Colin and Laeuger, Andrew and Li, Fulin and Lu, Neil and Nathan, Rowina S. and Ma, Sizheng and Chen, Yanbei and Sun, Ling},
  title         = {From Driven Oscillations to Free Ringdown: A Particle Plunging into {Kerr}},
  year          = {2026},
  eprint        = {2609.09531},
  archiveprefix = {arXiv},
  primaryclass  = {gr-qc},
  doi           = {10.48550/arXiv.2609.09531}
}

@misc{NakamotoEtAl2026PoleSplitting,
  author        = {Nakamoto, Nao and Oshita, Naritaka and Takeda, Hiroki},
  title         = {Theoretical Aspects of Direct Waves in {Kerr} Black Holes: Pole-Splitting Method for Ringdown Analysis},
  year          = {2026},
  eprint        = {2609.11908},
  archiveprefix = {arXiv},
  primaryclass  = {gr-qc},
  doi           = {10.48550/arXiv.2609.11908}
}

@article{Watarai2024,
  author  = {Watarai, D.},
  title   = {Ringdown of a Postinnermost Stable Circular Orbit of a Rapidly Spinning Black Hole: Mass Ratio Dependence of Higher Harmonic Quasinormal Mode Excitation},
  journal = {Physical Review D},
  volume  = {110},
  pages   = {124029},
  year    = {2024},
  doi     = {10.1103/PhysRevD.110.124029},
  eprint  = {2408.16747},
  archiveprefix = {arXiv},
  primaryclass  = {gr-qc}
}

@article{ReisswigPollney2011,
  author  = {Reisswig, Christian and Pollney, Denis},
  title   = {Notes on the Integration of Numerical Relativity Waveforms},
  journal = {Classical and Quantum Gravity},
  volume  = {28},
  pages   = {195015},
  year    = {2011},
  doi     = {10.1088/0264-9381/28/19/195015},
  eprint  = {1006.1632},
  archiveprefix = {arXiv},
  primaryclass  = {gr-qc}
}

@article{QuadraticGravity1,
    author = "Khera, Neev and Ma, Sizheng and Yang, Huan",
    title = "{Quadratic Mode Couplings in Rotating Black Holes and Their Detectability}",
    eprint = "2410.14529",
    archivePrefix = "arXiv",
    primaryClass = "gr-qc",
    doi = "10.1103/PhysRevLett.134.211404",
    journal = "Phys. Rev. Lett.",
    volume = "134",
    number = "21",
    pages = "211404",
    year = "2025"
}

@article{Ripley:Evolution,
    author = "Ripley, Justin L. and Loutrel, Nicholas and Giorgi, Elena and Pretorius, Frans",
    title = "{Numerical computation of second order vacuum perturbations of Kerr black holes}",
    eprint = "2010.00162",
    archivePrefix = "arXiv",
    primaryClass = "gr-qc",
    doi = "10.1103/PhysRevD.103.104018",
    journal = "Phys. Rev. D",
    volume = "103",
    pages = "104018",
    year = "2021"
}

@article{Anil:2007jw,
    author = "Zenginoglu, Anil",
    title = "{Hyperboloidal foliations and scri-fixing}",
    eprint = "0712.4333",
    archivePrefix = "arXiv",
    primaryClass = "gr-qc",
    reportNumber = "AEI-2007-177",
    doi = "10.1088/0264-9381/25/14/145002",
    journal = "Class. Quant. Grav.",
    volume = "25",
    pages = "145002",
    year = "2008"
}

@misc{Shi:2026mrr,
    author = "Shi, Jun-Xi and He, Zhen-Tao and Jiao, Jiageng and Lai, Jing-Qi and Shao, Caiying and Tian, Yu and Zhang, Hongbao",
    title = "{Can Oscillatory and Persistent Nonlinearities Be Bridged in Black Hole Ringdown?}",
    eprint = "2603.07469",
    archivePrefix = "arXiv",
    primaryClass = "gr-qc",
    month = "3",
    year = "2026"
}

@article{Zimmerman:2011dx,
    author = "Zimmerman, Aaron and Chen, Yanbei",
    title = "{New Generic Ringdown Frequencies at the Birth of a Kerr Black Hole}",
    eprint = "1106.0782",
    archivePrefix = "arXiv",
    primaryClass = "gr-qc",
    doi = "10.1103/PhysRevD.84.084012",
    journal = "Phys. Rev. D",
    volume = "84",
    pages = "084012",
    year = "2011"
}

@article{Shao:2023yjx,
    author = "Shao, Cai-Ying and Hu, Yu and Shao, Cheng-Gang",
    title = "{Parameter estimation for Einstein-dilaton-Gauss-Bonnet gravity with ringdown signals}",
    eprint = "2307.02084",
    archivePrefix = "arXiv",
    primaryClass = "gr-qc",
    doi = "10.1088/1674-1137/ace522",
    journal = "Chin. Phys. C",
    volume = "47",
    number = "10",
    pages = "105101",
    year = "2023"
}

@article{Mino:2008at,
    author = "Mino, Yasushi and Brink, Jeandrew",
    title = "{Gravitational Radiation from Plunging Orbits: Perturbative Study}",
    eprint = "0809.2814",
    archivePrefix = "arXiv",
    primaryClass = "gr-qc",
    doi = "10.1103/PhysRevD.78.124015",
    journal = "Phys. Rev. D",
    volume = "78",
    pages = "124015",
    year = "2008"
}

@article{sxs1,
   title={The SXS collaboration catalog of binary black hole simulations},
   volume={36},
   ISSN={1361-6382},
   url={http://dx.doi.org/10.1088/1361-6382/ab34e2},
   DOI={10.1088/1361-6382/ab34e2},
   number={19},
   journal={Classical and Quantum Gravity},
   publisher={IOP Publishing},
   author = {Boyle, Michael and others},
   year={2019},
   month=Sept, pages={195006} }

@article{sxs2,
   title={The SXS collaboration's third catalog of binary black hole simulations},
   volume={42},
   ISSN={1361-6382},
   url={http://dx.doi.org/10.1088/1361-6382/adfd34},
   DOI={10.1088/1361-6382/adfd34},
   number={19},
   journal={Classical and Quantum Gravity},
   publisher={IOP Publishing},
   author = {Scheel, Mark A. and others},
   year={2025},
   month=Oct, pages={195017} }

@misc{SXS3,
  author = {{SXS Collaboration}},
  title  = {{SXS Gravitational Waveform Database}},
  url    = {https://www.black-holes.org/waveforms},
  note   = {Accessed: 2025-12-20}
}

@article{EOB-waveform,
    author = "Boh{\'e}, Alejandro and others",
    title = "{Improved effective-one-body model of spinning, nonprecessing binary black holes for the era of gravitational-wave astrophysics with advanced detectors}",
    eprint = "1611.03703",
    archivePrefix = "arXiv",
    primaryClass = "gr-qc",
    reportNumber = "LIGO-P1600315",
    doi = "10.1103/PhysRevD.95.044028",
    journal = "Phys. Rev. D",
    volume = "95",
    number = "4",
    pages = "044028",
    year = "2017"
}

@misc{KuntzDellaRocca2026Redshift,
  author = {Kuntz, Adrien and Della Rocca, Matteo},
  title = {Vanishing of all redshift modes in {Schwarzschild} ringdown},
  year = {2026},
  eprint = {2606.07173},
  archivePrefix = {arXiv},
  primaryClass = {gr-qc},
  url = {https://arxiv.org/abs/2606.07173v1}
}

@misc{DW-lightring,
      title={Moving Beyond Quasinormal Modes and Reinterpreting Direct Waves in Binary Black Hole Merger-Ringdowns}, 
      author={Richard Dyer and Anuj Kankani and Gareth Arturo Marks and Sean T. McWilliams and Christopher J. Moore},
      year={2026},
      eprint={2610.25176},
      archivePrefix={arXiv},
      primaryClass={gr-qc},
      url={https://arxiv.org/abs/2610.25176}, 
}

@article{Nonlinear6,
    author = "He, Zhen-Tao and Du, Jia and Jiao, Jiageng and Shao, Caiying and Shi, Junxi and Tian, Yu and Zhang, Hongbao",
    title = "{Numerical computation of electromagnetically sourced nonlinear tails}",
    eprint = "2508.20499",
    archivePrefix = "arXiv",
    primaryClass = "gr-qc",
    doi = "10.1103/xrs3-82cb",
    journal = "Phys. Rev. D",
    volume = "112",
    number = "10",
    pages = "104008",
    year = "2025"
}

@Article{Teukolsky1973,
  author  = {Teukolsky, Saul A.},
  journal = {Astrophys. J.},
  title   = {{Perturbations of a rotating black hole. 1. Fundamental equations for gravitational electromagnetic and neutrino field perturbations}},
  year    = {1973},
  pages   = {635--647},
  volume  = {185},
  doi     = {10.1086/152444},
}

\end{document}